\documentclass[a4paper,UKenglish,cleveref, autoref, thm-restate]{lipics-v2021}
\usepackage{booktabs}
\usepackage{graphicx}
\usepackage{subcaption}

\newif\ifreviewmode
\newif\ifshowcomments
\newif\ifrevisionmode

\reviewmodefalse
\showcommentsfalse
\revisionmodefalse
\nolinenumbers

\ifreviewmode
  \newcommand{\df}[1]{\textcolor{blue}{#1}}
\else
  \newcommand{\df}[1]{#1}
\fi

\ifrevisionmode
  \newcommand{\rv}[1]{\textcolor{red}{#1}}
\else
  \newcommand{\rv}[1]{#1}
\fi

\ifshowcomments
  \newcommand{\sw}[1]{\textcolor{red}{\textbf{[SW:} #1\textbf{]}}}
\else
  \newcommand{\sw}[1]{}
\fi

\title{Contextual Associations Between Webpage Elements for Web Accessibility: An Empirical Study}

\author{Kishan Rakesh}{Department of Computer Science, University of Texas at Dallas, USA \and \url{https://kishanrakesh.github.io/portfolio/}} {Kishan.Rakesh@UTDallas.edu}{https://orcid.org/0009-0008-8722-7916}{}

\author{Shiyi Wei}{Department of Computer Science, University of Texas at Dallas, USA \and \url{https://personal.utdallas.edu/~swei/}}{swei@utdallas.edu}{https://orcid.org/0000-0002-2826-1857}{}

\authorrunning{K. Rakesh and S. Wei}

 \Copyright{Kishan Rakesh and Shiyi Wei}



\begin{document}

\maketitle

\begin{abstract}
\textbf{[Context]} Screen reader users navigating webpages by element list often encounter accessible names such as ``Read more'' that are valid under the W3C Accessible Name and Description Computation specification but uninterpretable in isolation. The surrounding elements that would make these names meaningful exist in the page but are not linked to the target by any mechanism. No prior work has empirically studied how to select which surrounding elements are contextually relevant to a given target. 
\textbf{[Objective]} This registered report investigates whether human-perceived contextual associations between webpage elements can be recovered from the accessibility tree using link prediction, and whether the learned associations generalize across websites. 
\df{\textbf{[Method]} We will construct a dataset of human-annotated contextual associations on 35 websites, stratified across the Tranco top-million list, with three independent annotators per page. Each page is represented as a graph derived from its accessibility tree, augmented with spatial and semantic features from the DOM and CSS. We compare four machine learning models (MLP, GCN, GAT, and SEAL) against two heuristic baselines under leave-one-site-out cross-validation with a pre-registered statistical framework, using Hit@K and MRR. 
\textbf{[Results]} We have conducted a five-site author-annotated pilot study to establish the pipelines and parameterize the power simulation, with pilot Hit@10 ranging from 0.16 to 0.85 across four learned models and 0.08 to 0.30 across two heuristic baselines. The final results will be reported after the planned experiments and analyses are completed.}
\textbf{[Conclusion]} The study contributes a human-annotated dataset of contextual associations on webpages, an empirical evaluation of link prediction for context selection on accessibility-tree graphs, and a cross-site generalization analysis.
\end{abstract}

\section{Introduction}
\label{sec:intro}

Web accessibility failures remain widespread despite legal regulations and guidelines. The WebAIM Million 2026 study \cite{webaimMillion2026} found that 95.9\% of the top one million home pages contain detectable Web Content Accessibility Guidelines (WCAG \cite{wcag22}) failures. Among the most consequential accessibility problems for screen reader users are poor accessible names, including missing, empty, or non-descriptive labels for interactive elements \cite{webaimMillion2026}. Screen reader users who navigate via element-list modes (such as VoiceOver Rotor \cite{appleVoiceOverRotor}, NVDA Elements List \cite{nvdaUserGuide}, JAWS virtual cursor \cite{jawsUserGuide}) receive these names as decontextualised strings without surrounding content; the accessible name is the only information available per element. The WebAIM Screen Reader Survey \#10 2024 \cite{webaimSurvey10_2024}, drawing on 1,539 respondents, reported non-descriptive link and button names among the most frequently cited problems.

An unresolved cause of uninformative accessible names is the absence of a computational model for context selection, i.e., determining which surrounding elements make a given name interpretable. \df{ACCNAME 1.1~\cite{accname11} does not infer visual contextual relationships across structural distance: a link labeled ``Read more'' passes the algorithm but derives its meaning from an article heading that may sit several DOM levels higher.}

\begin{figure}[t]
  \centering
  \begin{subfigure}[b]{0.50\textwidth}
    \centering
    \includegraphics[width=\textwidth]{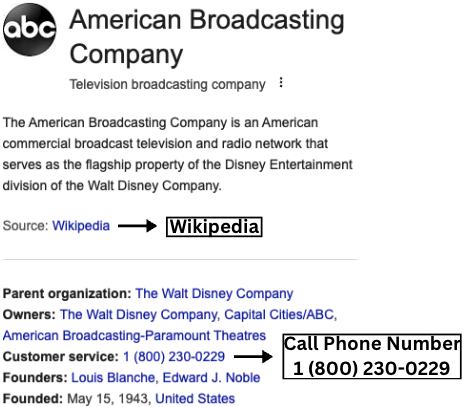}
    \caption{Google search panel and accessible names generated by VoiceOver Rotor.}
    \label{fig:original-panel}
  \end{subfigure}
  \hfill
  \begin{subfigure}[b]{0.45\textwidth}
    \centering
    \includegraphics[width=\textwidth]{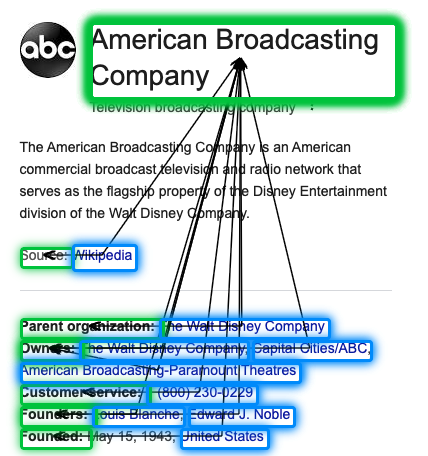}
    \caption{\df{Annotated panel showing contextual associations.}}
    \label{fig:annotated-panel}
  \end{subfigure}
  \caption{\df{The context-selection problem and our annotation interface, illustrated on a Google search knowledge panel.}}
  \label{fig:panel}
\end{figure}

\df{Figure~\ref{fig:original-panel} illustrates the problem on a Google knowledge panel for the American Broadcasting Company. The panel contains interactive links whose accessible names are ACCNAME-valid but uninterpretable in isolation. The link labeled ``Wikipedia'' sits next to the text ``Source:'' and beneath the heading ``American Broadcasting Company'', so a sighted user reads it as \emph{the Wikipedia page for ABC}. A screen-reader user navigating by link list hears only ``Wikipedia''. Similarly, the link ``1~(800)~230-0229'' is interpretable only because the adjacent text ``Customer service'' and the heading appear nearby; without them it is a bare phone number. In both cases the context that makes the link interpretable is visible to a sighted user but unreachable by ACCNAME. Recovering these associations automatically requires knowing which surrounding elements a sighted user would judge as contextually relevant.}

\df{Recent LLM-based accessibility tools have made progress on accessible-name repair but do not learn which surrounding elements provide the contextual signal a name needs. ACCESS~\cite{huang2024accesspromptengineeringautomated} corrects per-violation from the affected element alone; AccessGuru~\cite{10.1145/3663547.3746360} passes each violating element to an LLM with URL-level metadata but no signal about surrounding DOM elements. Context-aware approaches exist for image alt text. Gubbi Mohanbabu and Pavel~\cite{10.1145/3663548.3675658} combine fixed proximity, layout, and CLIP-based weights, and Dang et al.~\cite{10.1007/978-3-030-93736-2_20} train a graph network to select a single text node per image. However, neither learns multi-element contextual associations of the kind the ABC knowledge-panel example illustrates, and neither extends to accessible names for links, buttons, or form controls. To our knowledge, no prior work has empirically learned which surrounding elements provide the contextual information needed to interpret a given target element on a webpage.}

\df{We formulate context selection as link prediction between webpage elements: given a source element, predict which surrounding elements provide the contextual information that makes its accessible name interpretable. We collect human-judged contextual associations using a browser-based annotation tool, model each page as a graph derived from its accessibility tree augmented with spatial and semantic features from the DOM and CSS, and train models to recover the human-annotated edges. The output is a ranked retrieval list that a downstream LLM-based name generator can use as context.}

\df{This Stage-1 registered report is organized around four research questions.
\textbf{RQ1 (Annotation Analysis):} Do sighted users perceive consistent contextual associations between webpage elements, and what structural and spatial properties characterize these associations?
\textbf{RQ2 (Learnability):} Can learned models predict human-perceived contextual associations between webpage elements?
\textbf{RQ3 (Generalization):} Do learned contextual associations generalize across held-out websites?
\textbf{RQ4 (Feature Importance):} What is the relative contribution of graph structure, node feature groups, and pair feature groups to prediction quality?
RQ1 is answered first because it produces the ground truth on which RQ2--RQ4 depend: without evidence that humans agree on the task, supervised learning has no defensible target. If inter-annotator agreement is too low, the task may be too subjective to support learning, and that null outcome would itself be informative.}

\df{Pilot results suggest that an engineered-feature MLP, and potentially graph-aware models, will outperform heuristic baselines by at least $\delta=0.10$ in Hit@10; whether graph structure adds further value beyond engineered features is a secondary question. We pre-register $\delta=0.10$ as the threshold for meaningful improvement over the strongest heuristic baseline; failure to exceed it, or failure to generalize across LOSO folds, would constitute evidence that the accessibility tree and engineered features provide insufficient signal for context selection.}

\df{This study contributes: (i)~a human-annotated dataset of directed contextual associations across 35 websites with three annotators per page, stratified across the Tranco top-million list~\cite{tranco2019}; (ii)~an empirical evaluation of link prediction for context selection on accessibility-tree graphs; (iii)~a cross-site generalization analysis under LOSO cross-validation with a pre-registered statistical framework; and (iv)~data-capture, graph-construction, and annotation pipelines released as artifacts for replication.}

\section{Background}
\label{sec:background}

\textbf{Accessibility Tree (AX Tree) for Webpage Representation.}
The AX tree has been validated as a computationally effective webpage representation: WebArena \cite{zhou2024webarena} adopts it as the primary observation modality for autonomous web agents, and Mind2Web \cite{10.5555/3666122.3667342} shows that filtering HTML into accessibility-tree-like representations improves LLM performance on web interaction tasks. Google's on-device Reading Mode \cite{barcik2023readingmode} applies graph neural networks (GNN) to AX trees with spatial, textual, and role features for node classification. However, the W3C ACCNAME 1.1 \cite{accname11}, which AX node names are based on, derives names from individual elements and their immediate markup context, with no mechanism for incorporating inter-element relationships.

\textbf{Visual Relationships in Web UI Layouts.}
The divergence between visual structure and document hierarchy is well-established: Cai et al.~\cite{cai2003vips} observed in the VIPS algorithm that same-parent DOM nodes are not necessarily semantically related. \df{Human-annotated relationship datasets capture these groupings as directed edges at scale: RICO Semantics \cite{sunkara-etal-2022-towards} provides about 66,000 label--element associations in mobile UIs, and FUNSD \cite{8892998} provides entity-linking edges between form fields. Our dataset extends this line of work beyond specific relationships such as icon--label pairs to any elements that humans judge to provide context.}

\textbf{Graph-Based Approaches to UI Understanding.}
\df{Graph neural networks have been applied to UI and web understanding by Graph4GUI \cite{10.1145/3613904.3642822} and WebEmbed \cite{10.1007/978-3-031-93598-5_27}, with WebEmbed showing that GNN-learned embeddings generalise across websites. Dang et al.~\cite{10.1007/978-3-030-93736-2_20} provide the most relevant cross-site precedent, training GNNs on DOM-tree graphs to select relevant text nodes as image context, with evaluation explicitly split by website across hundreds of news sites. For the link prediction objective, SEAL \cite{10.5555/3327345.3327423}, GraphSAGE \cite{10.5555/3294771.3294869}, and GAT \cite{veličković2018graph} are established methods we adopt or compare against.}

\df{These works show that AX trees, UI-layout relationships, and graph-based webpage representations can support downstream learning. None recover the open-ended contextual associations a sighted user would make on a webpage; that is the gap this study addresses.}

\section{\df{Study Overview}}
\label{sec:overview}

\df{This report makes the dependency between sections explicit. Section~\ref{sec:study-design} specifies the dataset, graph construction, models, and evaluation setup used to answer RQ1--RQ4. \rv{Section~\ref{sec:prelim} reports a five-site pilot providing feasibility validation and the Hit@10 estimates that parameterize the power simulation. Section~\ref{sec:planned} contains the pre-registered analyses and statistical framework.} Sections~\ref{sec:threats} and~\ref{sec:ethics} cover validity and ethics.}

\section{Study Design}
\label{sec:study-design}
\subsection{Dataset Creation}

\textbf{Site Selection.}
\df{Sites are sampled from the Tranco top-million list~\cite{tranco2019} using stratified sampling across seven log-scaled traffic bands from rank 1--10 up to rank 300{,}001--1{,}000{,}000. Five sites are sampled per band, yielding $N=35$ total. Stratification spans the traffic distribution a typical user encounters rather than only the highest-traffic sites\rv{; bands are log-scaled because traffic is concentrated at the top, so linear bins would give the low-traffic tail equal weight to the high-traffic head despite its far smaller reach}. Within each band, sites are drawn uniformly at random from Tranco entries that pass the inclusion criteria below. We do not enforce within-band category or layout constraints, since existing webpage taxonomies categorize sites by content topic rather than by the layout properties relevant to our task, offering no clear basis for balanced layout coverage.}

\df{We commit to $N=35$ based on a power simulation parameterized by the per-site pilot Hit@10 differences (Table~\ref{tab:pilot-dataset}). The primary registered test (Section~\ref{sec:planned}, RQ3) compares the MLP against the strongest heuristic baseline; the simulation uses the pilot MLP-vs-tree-distance per-site differences as the $\sigma_d$ estimate ($\sigma_d=0.174$), as tree distance was the stronger pilot heuristic. At $N=35$, this yields 91\% power to detect $\delta=0.10$ and 98\% at $\delta=0.12$; the pilot-observed mean difference ($0.340$) is detected at $>99\%$ power. A sensitivity analysis with $\sigma_d$ inflated by $1.5\times$ retains 91\% power for $\delta=0.15$. Power was estimated by Monte Carlo simulation using the same two-tailed sign-flip permutation test specified in Section~\ref{sec:planned}: 10{,}000 datasets per $(N, \delta)$ cell, power equal to the rejection rate at $\alpha=0.05$. \rv{The registered tests pool all 35 sites: the primary within-site paired test removes between-site variation, so the sign-flip null is exchangeable across sites.}}

\df{Each candidate page is screened against three inclusion criteria. A source candidate is a non-ignored AX node whose role belongs to the interactive-widget set (links, buttons, form controls), mirroring the elements surfaced by VoiceOver Rotor and NVDA Elements List. (C1)~The rendered viewport contains $\geq 50$ source-candidate AX nodes. The threshold supports per-page ranking estimates while keeping annotation effort near 30 minutes; in the pilot, bootstrap SD of mean F1@10 was roughly 0.012 at 50 sources. (C2)~The static-snapshot representation is stable (Data Capture below). (C3)~The page is in English and authentication, if any, can be legitimately satisfied. If fewer than five sites in a band qualify, replacement sampling within the band continues until five pass or the band is exhausted. Only band 1 (10 Tranco entries) is small enough for exhaustion to be a real concern; any shortfall is absorbed by sampling additional qualifying sites from the next band, cascading further if necessary, to preserve $N=35$. We report, in the Stage-2 paper, per-band acceptance counts and rejection counts broken out by failed criterion (C1, C2, C3), together with the site-category distribution of the final 35 sites.}

\textbf{Data Capture.}
\df{For each webpage, three artifacts are captured using a Playwright-based pipeline~\cite{playwright}. The pipeline produces (1)~a static HTML snapshot with scripts removed, (2)~a raw AX Tree extracted via Chrome DevTools Protocol~\cite{chromeDevToolsProtocol} containing AX nodes with roles, names, and states, and (3)~an enriched AX Tree augmented with CSS styles, bounding boxes, and XPaths for mapping elements to the underlying HTML. For pages requiring authentication, the pipeline supports an interactive capture mode that pauses for manual authentication before extraction.}

\df{For C2, the pipeline captures the initial viewport, the page bottom, and the initial viewport again. Source-candidate nodes are matched by XPath between the first and final top snapshots; unmatched nodes or nodes whose role/name changes are assigned IoU 0. Page stability is defined as the 5th-percentile IoU across source candidates. Before annotation begins, the IoU exclusion threshold is calibrated on 20 candidate pages sampled separately from the registered 35, using manual stable/unstable labels and a fixed threshold grid, maximizing balanced accuracy with ties broken toward the stricter threshold. The calibration set, selected threshold, and per-band C2 rejection counts will be reported in Stage 2.}

\textbf{Graph Construction.}
\df{A graph construction pipeline converts each enriched AX Tree into a graph. AX nodes that map to no on-page element are dropped; parent-child pairs with identical visible text are collapsed to the topmost node to reduce redundancy. The resulting graph contains far fewer nodes than the raw DOM while retaining semantic structure and inter-element relationships.}

\df{Each node carries a 34-dimensional feature vector with three groups\rv{, grounded in the proximity, salience, role, and structure cues identified in prior UI-understanding work~\cite{cai2003vips,barcik2023readingmode,10.1145/3663548.3675658}; the groups are defined so that RQ4 can attribute predictive contribution to each}:}
\df{\begin{itemize}
  \item \textbf{Spatial (8):} viewport-normalized $x$, $y$, width, height, area, aspect ratio; tree depth (depth-normalized); sibling index (parent-normalized).
  \item \textbf{Semantic (6+12):} encoded WAI-ARIA role ID, HTML tag ID, log-scaled visible-text length, child-node count, heading flag, text-node flag~\cite{waiAria12}; plus 12 features storing collapsed-subtree tag-frequency counts (e.g., \texttt{<div>}, \texttt{<span>}, \texttt{<a>}).
  \item \textbf{Style (8):} font size, font weight, opacity, foreground/background luminance, WCAG contrast ratio, focusable flag, visible flag.
\end{itemize}
Role IDs and tag IDs map to a vocabulary built from training-fold sites only; normalization statistics are likewise computed on training folds to prevent leakage.}

The edge set contains two types of edges: structural edges and ARIA relationship edges. The structural edge set consists of two sets of bidirectional relationships: parent-child and previous-next sibling. This preserves the AX tree structure while allowing message passing across sibling and child-parent element pairs. The ARIA relationship edges extract semantic relationships directly from HTML WAI-ARIA attributes, such as \texttt{aria-labelledby}, \texttt{aria-describedby}, \texttt{aria-controls}, and \texttt{aria-owns}~\cite{waiAria12}. These relationships are sparse across webpages but encode semantic associations where present.

\df{\textbf{Dataset Preparation.}
The graph-construction pipeline emits a per-site graph $G_s=(V_s,E_s,X_s)$ from the enriched AX tree; annotation records are projected onto $G_s$ by resolving each XPath to its AX node, yielding a per-site positive edge set $P_s$. For training, negatives are sampled from $\{(u,v)\mid u,v\in V_s,\,(u,v)\notin P_s\}$ with $u$ restricted to source-candidate nodes, at a 1:1 ratio with positives. All artifacts (graphs, features, annotation records, train/evaluation splits, model checkpoints, and evaluation outputs) will be released with the Stage-2 paper.}

\textbf{Annotation Protocol.}
\df{Annotators are recruited from UT~Dallas undergraduate and graduate students through an open volunteering call. Three unique annotators are assigned per page, yielding 105 annotation slots across the 35 sites. Based on pilot timing, each annotator takes 15--30 minutes per page after a one-time 30-minute interface tutorial; a volunteer who annotates 10 pages spends roughly 3--6 hours total. We anticipate a pool of 8--12 annotators, with different pages annotated by different annotator subsets.}

\df{Each annotator receives an annotation-instructions document containing: (i)~a functional definition, \textit{an element is contextually relevant to a source if it provides information that helps a sighted user understand the source's purpose or meaning}; (ii)~3--4 worked examples drawn from the pilot, and (iii)~a step-by-step tutorial walkthrough of the annotation interface. Annotators are not given a typology of relationship classes; the relation is an untyped directed association, which avoids biasing judgments toward pre-specified categories. Annotators may leave a source unmarked when no surrounding element appears contextually relevant.}

\df{Annotation is collected through a custom JavaScript-overlay interface on the rendered page snapshot. The annotator interacts visually by clicking on rendered elements on the webpage and each click is recorded as an XPath that maps deterministically to an AX node in the captured tree. The interface displays each created relationship as an arrow overlaid on the rendered page so the annotator can verify and revise.}

\df{The union of edges across the three annotators per page is used as the positive label set for model training and primary evaluation. A high-agreement subset, consisting of edges selected by at least two of three annotators, is used as a secondary evaluation set. Inter-annotator agreement is reported per-page as mean pairwise F1 over the directed edge sets, treating one annotator's set as predictions and another's as ground truth and averaging across the three pairs. F1 is the primary agreement metric because the task is set retrieval over a highly imbalanced edge space: Cohen's and Fleiss's $\kappa$ would be dominated by trivial agreement on the large negative class, making them uninformative for this setting.}

\subsection{Link Prediction Models}
\df{We evaluate two heuristic baselines and four learned models. The heuristic baselines rank candidates by Euclidean distance to the source (spatial proximity) and by shortest-path distance in the AX tree (tree distance). The four learned models share a common input projection that lifts the 34-dimensional node feature vector to a hidden dimension, and a common edge-prediction head that concatenates the source and target node representations with an 11-dimensional pair feature vector (Euclidean distance, bounding-box overlap metrics, relative-position quadrant, LCA depth, and LCA-to-endpoint hop distances) and passes the result through a two-layer MLP with sigmoid output. Models differ in how node representations are computed.}

\df{The MLP baseline applies the input projection and an MLP encoder over node features alone, with no message passing, distinguishing the contribution of graph structure from that of node and pair features. The GCN~\cite{kipf2017semisupervised} stacks uniform-aggregation message-passing layers over the structural and ARIA edges; it is the simplest graph model. The GAT~\cite{veličković2018graph} replaces uniform aggregation with multi-head attention, allowing it to weight neighbor contributions differently for each pair. SEAL~\cite{10.5555/3327345.3327423} extracts a fixed-radius enclosing subgraph around each candidate pair, applies double-radius node labeling, and runs a graph-level classifier over the labeled subgraph, enabling local-structure reasoning without global message passing.}

\df{All learned models use binary cross-entropy loss and the Adam optimizer~\cite{kingma2015adam}. Training pairs are sampled at a 1:1 negative-to-positive ratio, following common practice in graph-based link-prediction benchmarks~\cite{10.5555/3327345.3327423,kipf2016vgae,hu2020ogb}; evaluation is performed against the full candidate pool, so Hit@K and MRR are computed under realistic class imbalance. Within each LOSO fold, training uses the union annotation set from training sites; the held-out site is reserved for evaluation.}

\df{Hyperparameters are selected per fold per model by random search over 50 trials, with the trial maximizing validation Hit@10 chosen. The shared search space covers hidden dimension $\in \{64, 128, 256\}$, depth $\in \{2, 3, 4\}$, dropout $\in \{0.1, 0.3, 0.5\}$, learning rate $\in \{10^{-4}, 10^{-3}, 10^{-2}\}$, and weight decay $\in \{0, 10^{-5}, 10^{-4}\}$. Model-specific extensions are: attention heads $\in \{2, 4, 6\}$ for GAT and enclosing-subgraph hop count $\in \{1, 2, 3\}$ for SEAL. The search budget and grid are fixed before annotation data collection begins.}

\subsection{Evaluation Setup}
\df{For each interactive source, the candidate target pool comprises all other non-ignored AX nodes in the page graph (including headings, body text, labels, and group titles, not only other interactive elements), ranging from approximately 500 to 1{,}900 nodes per page in the pilot. Because annotated targets are sparse against this candidate pool (typically fewer than 10 per source), we use ranking metrics rather than classification metrics. The primary metric is Hit@K~\cite{JMLR:v10:gunawardana09a}: for a source with one or more annotated targets, Hit@K is 1 if any annotated target appears in the top-$K$ predictions and 0 otherwise, reflecting the practical scenario of providing the top $K$ predicted contextual elements to a downstream LLM-based name generator. We assess $K \in \{10, 5, 1\}$. The secondary metric is mean reciprocal rank (MRR)~\cite{JMLR:v10:gunawardana09a}, the reciprocal rank of the highest-ranked annotated target per source, averaged across sources; Precision@10, Recall@10, PR-AUC, and F1 are reported as additional metrics. Evaluation uses leave-one-site-out (LOSO) cross-validation with one fold per site; the held-out site is the test set, with one of the remaining sites held out for validation. Statistical testing of the per-site Hit@10 results is specified in Section~\ref{sec:planned}.}

\section{\df{Pilot Study}}
\label{sec:prelim}

\df{We conducted a pilot study on five websites (Amazon, Facebook, Google, Twitter/X, Wikipedia) to validate that the full data and annotation pipeline produces the artifacts the registered study requires. The pilot does not answer any registered research question; it functions as feasibility validation as described in Section~\ref{sec:overview}. The pipelines produced the expected artifacts and the annotation tool recorded source and target XPaths correctly against the captured snapshot. We then trained all four learned models and both heuristic baselines on the author-annotated data to obtain per-method Hit@10 estimates for the power simulation. Dataset statistics and pilot Hit@10 per site for all six methods are reported in Table~\ref{tab:pilot-dataset}.}

\begin{table}[t]
\centering
\caption{Pilot dataset summary and \df{pilot Hit@10 by method.}}
\label{tab:pilot-dataset}
\small
\df{\begin{tabular}{lrrr|rrrr|rr}
\toprule
 & \multicolumn{3}{c|}{Dataset} & \multicolumn{4}{c|}{Learned} & \multicolumn{2}{c}{Heuristic} \\
Site & Nodes & Edges & Ann. & MLP & GCN & GAT & SEAL & Spatial & Tree \\
\midrule
Amazon    & 1893 & 4022 & 608 & 0.496 & 0.436 & 0.354 & 0.535 & 0.227 & 0.300 \\
Facebook  & 1011 & 1636 & 405 & 0.575 & 0.283 & 0.300 & 0.442 & 0.117 & 0.075 \\
Google    & 1860 & 3032 & 549 & 0.451 & 0.362 & 0.160 & 0.531 & 0.230 & 0.221 \\
Twitter/X & 514  & 1636 &  79 & 0.435 & 0.826 & 0.826 & 0.848 & 0.109 & 0.217 \\
Wikipedia & 1536 & 3695 & 277 & 0.726 & 0.603 & 0.536 & 0.480 & 0.156 & 0.168 \\
\bottomrule
\end{tabular}}
\end{table}

\df{The pilot Hit@10 estimates give the variance parameter ($\sigma_d = 0.174$) that the power simulation in Section~\ref{sec:study-design} uses to fix the registered sample size $N=35$. They are not powered to support inference about RQ1--RQ4, because the pilot was author-annotated, trained on all five sites combined into a single training set (a pooled setting, in contrast to the LOSO cross-validation used in the registered study), and run without the hyperparameter search.}

\section{Planned Analyses}
\label{sec:planned}

\df{The registered dataset covers 35 websites with three independent annotators per page. The analyses below follow one paragraph per RQ. The primary evaluation metric throughout is Hit@10; MRR and Hit@5/Hit@1 are reported as secondary metrics.}

\textbf{RQ1: Annotation Analysis.}
\df{We report the distribution of per-page F1 (defined in the Annotation Protocol) across all 35 pages, and additionally report Krippendorff's $\alpha$ with Jaccard distance~\cite{hayes2007krippendorff} as a chance-corrected supplement. RQ1 is treated as exploratory: we do not pre-register a pass/fail agreement threshold, because no existing literature establishes one for open-ended relation annotation on webpage elements. Model evaluation in RQ2--RQ4 is reported on both the union annotation set and the high-agreement subset (defined in the Annotation Protocol) in parallel, so that conclusions do not depend on which label set is used. We then analyze the structural properties of the annotated edges (accessibility-tree hop distance, lowest-common-ancestor depth, Euclidean bounding-box distance, and role-pair frequencies), comparing annotated versus non-annotated pairs on the same page, and comparing the union set against the high-agreement subset to assess whether high-agreement edges are structurally distinguishable from single-selected annotations.}

\textbf{RQ2: Learnability.}
We compare four learned models (MLP, GCN, GAT, SEAL) against two heuristic baselines: spatial-proximity ranking and tree-distance ranking. For each model and each baseline, we compute Hit@10, Hit@5, Hit@1, and MRR in the LOSO folds. We evaluate on the union annotation set and on the high-agreement subset defined in the Annotation Protocol. \df{Statistical comparison of learned models against the strongest heuristic baseline uses the framework specified for RQ3.}

\textbf{RQ3: Generalization.}
LOSO cross-validation produces one test fold per site. We report per-site Hit@10 and MRR for each learned model and each baseline. \df{The primary registered test is a two-tailed paired permutation test on per-site Hit@10 comparing the MLP against the strongest heuristic baseline, approximating the null distribution by Monte Carlo sign-flipping across sites (10{,}000 random sign assignments). As secondary tests, we run paired permutation tests comparing each of GCN, GAT, and SEAL against the strongest heuristic baseline, with Holm-Bonferroni correction across the three secondary comparisons. The strongest heuristic baseline for each comparison is selected per-fold as whichever of spatial-proximity or tree-distance has higher mean Hit@10 on the training sites in that fold. We report bootstrap 95\% confidence intervals on per-method Hit@10 by resampling sites with replacement (10{,}000 resamples). As a supplementary analysis, we apply Bayesian signed-rank tests~\cite{benavoli2017bayesian} to all four learned-vs-strongest-heuristic per-site Hit@10 differences, reporting posterior probabilities that each learned model exceeds the strongest heuristic baseline.}

\textbf{RQ4: Feature Importance.}
We take the best-performing graph model from RQ2 as the base and retrain it with one feature group removed at a time. The ablated groups are: (1)~spatial node features, (2)~semantic node features, (3)~style node features, (4)~spatial pair features, (5)~structural pair features, and (6)~ARIA relationship edges. For each ablation we report Hit@K metrics for the ablated model alongside the metrics for the full model. We rank the feature groups by the size of the drop and treat the ranking as exploratory.

\section{Threats to Validity}
\label{sec:threats}
We specify two forms of threats to validity: internal and external.

\textbf{Internal validity.} The annotation task captures contextual relevance as an untyped directed relation. This allows users to judge relevance \df{freely, but it may conflate different association types} such as heading--content, grouping, and semantic association. Annotations are not exhaustive, so some related pairs may remain unlabeled and be treated as negatives during training and evaluation. A further concern is annotation error: annotators may click the wrong element or miss intended associations. We mitigate this through a training exercise before annotation begins and by parallel reporting on the union and high-agreement label sets. \df{The annotator pool of 8--12 introduces inter-page variance in annotation style; we report inter-annotator agreement per-page rather than pooled so any variance is visible.}

\textbf{External validity.} Each site is represented by a single static snapshot. \df{\rv{The C2 stability check (Section~\ref{sec:study-design}) screens out pages whose layout depends on scroll-triggered DOM changes; modals, lazy-loaded content, and interaction-driven states remain out of scope.} A high C2 rejection rate would indicate that the static-snapshot method covers only a subset of the top million; we report per-band C2 rejection counts as a measure of this coverage limit. Stratified Tranco sampling broadens coverage but limits findings to the top million; sites outside the top million and non-English sites are out of scope, and long-tail generalization is future work. Within-band sampling does not enforce category or layout diversity, so the category distribution depends on the Tranco websites that pass screening. \rv{The category distribution is descriptive rather than targeted; $N=35$ is too small to measure layout-variation coverage, and we make no representativeness claim.} LOSO cross-validation addresses within-population generalization.}

\section{Ethical Considerations}
\label{sec:ethics}

Annotators will provide informed consent before participation, with the right to withdraw at any time. Annotation records contain only source and target XPaths, a study-specific annotator ID, and timestamps, and no direct personal identifiers. For authentication-required sites, captured snapshots may contain profile-specific or user-generated content, so any released artifacts will be reviewed and redacted as needed.

\section{Conclusions}
\label{sec:conclusions}
\df{This registered report formulates context selection for accessible name repair as link prediction on accessibility-tree graphs. We build a human-annotated dataset across 35 websites and pre-register an evaluation of learned models against heuristic baselines, designed so the dataset and structural analysis remain valuable even if the models do not beat the baselines. On acceptance, we will run the analyses as specified. The Stage-2 paper will include a dedicated deviations section reporting any changes from this protocol, their rationale, and their impact on interpretation.}

\bibliography{refs}

\end{document}